\documentclass[useAMS,usenatbib]{mn2e}
\usepackage{amsmath}
\usepackage{amssymb}
\usepackage{graphicx}
\usepackage{color}

\title[Feedback-limited Variable Accretion on to Planets]
  {Feedback-limited Accretion:  Variable Luminosity from Growing Planets}
\author[M. G\'arate et al.]
  {M.~G\'arate,$^{1,2,8}$\thanks{Email: garate@mpia.de }
  J.~Cuadra,$^{1,3,4,5}$ M.~Montesinos,$^{6,7,5}$ P.~Ar\'evalo$^{6}$
             \\
  $^1$Instituto de de Astrof\'isica, Pontificia Universidad Cat\'olica de Chile, Santiago, Chile\\
  $^2$University Observatory, Faculty of Physics, Ludwig-Maximilians-Universit\"at M\"unchen, Scheinerstr.\ 1, 81679 Munich, Germany\\
  $^3$Max-Planck-Institut f\"ur extraterrestriche Physik (MPE), D-85748 Garching, Germany\\
  $^4$Departamento de Ciencias, Facultad de Artes Liberales, Universidad Adolfo Ib\'a\~nez, Av.\ Padre Hurtado 750, Vi\~na del Mar, Chile \\
  $^5$N\'ucleo Milenio de Formaci\'on Planetaria (NPF), Chile \\
  $^6$Instituto de F\'isica y Astronom\'ia, Universidad de Valpara\'iso, Av.\ Gran Breta\~na 1111,  Casilla 5030, Valpara\'iso, Chile   \\ 
  $^7$South America Center for Astronomy, National Astronomical Observatories, CAS, 
Beijing 100012, China\\
  $^8$Max-Planck-Institut f\"ur Astronomie, K\"onigstuhl 17, 69117, Heidelberg, Germany}
\date{\today}

\pagerange{\pageref{firstpage}--\pageref{lastpage}} \pubyear{2020}

\def\LaTeX{L\kern-.36em\raise.3ex\hbox{a}\kern-.15em
    T\kern-.1667em\lower.7ex\hbox{E}\kern-.125emX}

\def\simlt{\lower.5ex\hbox{$\; \buildrel < \over \sim \;$}}
\def\simgt{\lower.5ex\hbox{$\; \buildrel > \over \sim \;$}}

\begin{document}

\label{firstpage}

\maketitle

\begin{abstract}
Planets form in discs of gas and dust around stars, and continue to grow by accretion of disc material while available.  Massive planets clear a gap in their  protoplanetary disc, but can still accrete gas through a circumplanetary disk. For high enough accretion rates the planet should be detectable at infrared wavelengths.
As the energy of the gas accreted on to the planet is released, the planet surroundings heat up in a feedback process. 
We aim to test how this planet feedback affects the gas in the coorbital region and the accretion rate itself.
%
%
%
We modified the 2D code FARGO-AD to include a prescription for the accretion and feedback luminosity of the planet and use it to model giant planets on 10 au circular and eccentric orbits around a solar mass star.
%
{We find that this feedback reduces but does not halt the accretion on to the planet, although this result might depend on the near-coincident radial ranges where both recipes are implemented.}
%
Our simulations also show that the planet heating gives the accretion rate a stochastic variability with an amplitude $\Delta \dot{M}_p \sim 0.1 \dot{M}_p$.
A planet on an eccentric orbit ($e=0.1)$ presents a similar variability amplitude, but concentrated on a well-defined periodicity of half the orbital period and weaker broadband noise, potentially allowing observations to discriminate between both cases.
Finally, we find that the heating of the coorbital region by the planet feedback alters the gas dynamics, reducing the difference between its orbital velocity and the Keplerian motion at the edge of the gap, which can have important consequences for the formation of dust rings.
\end{abstract}


\section{Introduction}
Protoplanetary discs are made of gas and dust orbiting a young central star. Giant planets like Jupiter form in these discs once a solid core is massive enough to capture the surrounding gas \citep{pollack96}, or if the gas itself becomes self-gravitating and fragments into clumps \citep{boss97}. 
Either way, after a massive planet like Jupiter has formed, it will produce perturbations in the protoplanetary disc due to its gravity, which end up opening a gap when the planet torque exceeds the viscous torque \citep{lin93,bryden99}, and forming a circumplanetary disc (CPD) within the Hill radius \citep{ayliffe09a}.
These planet--disc interactions define the early evolution of young planets, so understanding them is key to interpret the observational signatures they produce, such as the width of the gaps carved on the gas or dust distribution, the presence of spiral arms, or perturbations on the velocity field \citep[e.g.,][]{kanagawa15, dong15a, dong15b, perez15, dipierro16, pinte18, pinte19, teague18}.

In addition to those mostly gravitational effects, the accretion of gas on to planets provides another kind of planet formation signatures. After a gap has formed, the CPD will keep providing material to the newly formed planet, and its properties will determine the planet accretion rate \citep{rivier12, szulagyi14}.
As the planet accretes, the gravitational energy of the gas will be released, heating up the circumplanetary region \citep{klahr06, montesinos15}. 
During the last decade multiple ``hot spots'' have been detected within protoplanetary discs \citep[e.g.,][]{quanz13, reggiani14, sallum15}, culminating on the first confirmed discovery of a forming planet around the T Tauri star PDS 70 \citep{keppler18, mueller18}.
These detections can be used  to constrain the properties of the embedded planets once their interplay with the accreting gas is understood.

Besides the observable signatures, the accretion of material could also alter the dynamics of the system as the circumplanetary region is heated.
The heating produced by accretion on to a rocky core can change the torque exerted on it, slowing and even reversing the inward migration \citep{benitez15, masset17, guilera19}.  
For more massive planets formed in a disc fragmentation process, feedback limits their growth and alters their migration \citep{nayakshin13, stamatellos15, stamatellos18}, besides preventing further fragmentation \citep{mercer17}\footnote{See also \cite{nayakshin07} for earlier models on the scale of galactic nuclei.}.  On smaller scales, the CPD region can be partially depleted by the effect of the feedback \citep{montesinos15}, and the heating of the CPD can extend its vertical structure, turning it into an envelope \citep{szulagyi16, szulagyi17_Feedback}.

The goal of this work is to study the interaction between the luminosity of the planet and its accretion rate when both variables are coupled. This extends our previous work presented by \cite{montesinos15}, in which accretion was not taken into account, and the planet luminosity was a free parameter. 
Specifically, we want to study whether the planet heating hinders or even prevents the accretion process itself.  Additionally, we want to study the  variability that feedback induces on the accretion rate, and test if such variations would be observable and distinguishable from those due to an eccentric orbit \citep{dunhill15}.



A similar study was already conducted by \cite{klahr06}. Our approach is simpler, since we are using an energy transport equation instead of the flux-limited diffusion approximation, and we limit ourselves to two dimensions. However, our model gives us the possibility to explore a larger parameter space for longer times and at higher resolutions.

This paper is structured as follows.  In section \ref{sec_model} we present our physical model and numerical setup. In section \ref{sec_AccretionVariability} we present the results on the accretion variability, and in section \ref{sec_HeatingGap} we show the effects of the feedback on the gap structure. Then, in section \ref{sec_Discuss} we discuss the applications and limitations of our model, concluding in section \ref{sec_Summary}.

\section{Physical Model} \label{sec_model}

We use the grid-based hydrodynamics code FARGO-AD \citep{baruteau08} to simulate a gas disc around a solar-mass star, with a Jupiter-mass planet located at $r_\textrm{p}=10\,$au from the star.
Our simulations include the modifications described in \cite{montesinos15} that consider the planet luminosity and radiative cooling.
Additionally, we compute the expected accretion on to the planet, which is used to calculate the planet accretion luminosity. This energy is distributed over the CPD as a heating term in the energy equation. In the following subsections we describe the equations used to quantify each of these ingredients.
\subsection{Energy Equation}
The energy surface density $e$ is evolved using the equation described in \cite{dangelo03}:
\begin{equation}\label{EqEnergy}
  \frac{\partial e}{\partial t} +   \overrightarrow{\nabla} \cdot(e \overrightarrow{v} ) = -P \overrightarrow{\nabla}\cdot \overrightarrow{v} + Q_\nu^+ + Q_\textrm{p}^+ - Q^-,
\end{equation}
where $\overrightarrow{v}$ is the velocity, $P$ is the vertically integrated pressure, $Q_\nu^+$ and $Q_\textrm{p}^+$ are the viscous and planet heating terms, and $Q^-$ is the radiative cooling term. 
As described in \cite{montesinos15}, the system is closed with an ideal equation of state for the pressure as a function of mass surface density $\Sigma$ and mid-plane temperature $T$:
\begin{equation}\label{EqIdealGas}
 P= \Sigma  T \overline{R},
\end{equation}
and the energy density is related to the temperature using:
\begin{equation}\label{EqIdealEnergy}
 e= \Sigma  T \frac{\overline{R}}{\gamma -1 },
\end{equation}
with the specific gas constant $\overline{R}= k_\textrm{b}/\mu m_\textrm{H}$ and $k_\textrm{b}$ the Boltzmann constant. The adiabatic index is $\gamma=1.4$ and the mean molecular weight is $\mu=2.35$.

We do not take into account the heating by stellar irradiation, since for our parameters the viscosity and the planet feedback are the dominant heating sources.
%
\subsection{Viscosity}
For the viscosity due to shear in the disc we use the \cite{shakura73} prescription: 
\begin{equation}\label{EqViscosity}
 \nu=\alpha c_{\rm s}^{2}/\Omega_\textrm{K},
\end{equation}
where $c_{\rm s}$ is the sound speed, $\Omega_\textrm{K}$ is the Keplerian angular velocity, and $\alpha$ is a free parameter that we set to $\alpha=0.004$.

The energy dissipation due to viscosity contributes to heating the disc through the following term,
\begin{equation}
 Q_{\nu}^+= \frac{1}{2\nu\Sigma}(\tau^2_{r,r}+2\tau^2_{r,\theta}+\tau^2_{\theta,\theta})+ \frac{2\nu\Sigma}{9}\nabla^2\overrightarrow{v},
 \end{equation}
which is described in \cite{dangelo03}, where $\tau_{r,r}$, $\tau_{r,\theta}$ and $\tau_{\theta,\theta}$ are the elements of the stress tensor.\\
\subsection{Accretion Model}\label{sec_AccModel}
We use the accretion model implemented in FARGO \citep{masset00}, which removes gas from the surface density within the Hill radius, using the following radial behaviour   \citep{kley99, duermann17},
\begin{equation}\label{EqAcc}
 \frac{d\Sigma_{\rm{acc}}}{dt}(r) = 
    \begin{cases}
	    -\Sigma(r)/t_{\rm{acc}}            &    	r< 0.45 R_{\rm{Hill}} \\
        -\frac{1}{3} \Sigma(r)/t_{\rm{acc}}  &	0.45R_{\rm{Hill}}<r< 0.75 R_{\rm{Hill}} \\
        0		&	 r>0.75R_{\rm{Hill}},
  \end{cases}
\end{equation}
where $r= |\overrightarrow r- \overrightarrow r_{\rm{p}}|$, the Hill radius is $R_{\rm{Hill}} = r_\textrm{p} \sqrt[3]{q/3}$, with $q$ the planet-to-star mass ratio, and the accretion timescale $t_{\rm{acc}}$ is a free parameter that corresponds to the time that it would take to accrete the mass of the circumplanetary disk.
The accretion rate on to the planet is equal to the removed surface density integrated over the disc,
\begin{equation}\label{EqAccPlanet}
 \dot{M}_\textrm{p} = \int -\frac{d\Sigma_{\rm{acc}}(r)}{dt} dA.
\end{equation}
In our simulation setup, however, we do not add the accreted mass $\dot{M}_\textrm{p} dt$ to the planet mass, as the final systems would not be dynamically equivalent and therefore not comparable in the posterior analysis.  We only use $\dot{M}_\textrm{p}$ to compute the heating that the accretion process produces.

Other works \citep{gressel13} have included a cooling recipe to follow the energy losses associated to this accretion process. In our work, we absorb additional cooling mechanisms in the efficiency factor $\epsilon$, described in the following section.
\subsection{Planet Heating}\label{sec_PlanetHeat}
The planet heating term $Q_\textrm{p}^+$ is calculated as
\begin{equation}\label{EqQheat}
 Q_\textrm{p}^+= \epsilon f(r) L_\textrm{p}(t),
\end{equation}
where $L_\textrm{p}$ is the total energy released by the planet, $f(r)$ is a smoothing function that distributes the energy within the CPD radius\footnote{With $\int_0 ^{R_\textrm{CPD}} 2 \pi r f(r) dr \approx 1$.} $R_\textrm{CPD}$, and $\epsilon$ is a free parameter efficiency factor that can be interpreted as the fraction of the energy emitted by the planet that is actually absorbed by the gas; its exact value will depend on the detailed physics of the accretion shock at the planet surface, and on the planet evolutionary stage \citep{marleau17, mordasini17}. For $R_\textrm{CPD}$ we follow \cite{crida09}  and consider $R_\textrm{CPD} = 0.6 R_{\rm{Hill}}$.
This work uses the same smoothing function described by \cite{montesinos15}:
\begin{equation}\label{f_smoot}
 f(r) = \begin{cases}
	   \frac{1}{\pi R_\textrm{CPD}^2} 5 \exp(-5 \frac{r^2}{R_\textrm{CPD}^2})            &    	r< R_\textrm{CPD} \\
       0		&	r  >R_\textrm{CPD}.     
      \end{cases}
\end{equation}

In our model the planet luminosity is calculated as
\begin{equation}\label{EqPLum}
 L_\textrm{p}(t)= \frac{1}{2} \frac{G M_\textrm{p}}{R_\textrm{p}}\dot{M_\textrm{p}}(t),
\end{equation}
considering the change of energy of a test particle in Keplerian motion when accreted at the planet radius $R_\textrm{p}$.  For the latter, we take the radius of Jupiter.\footnote{Notice that because the gas is accreted from a finite distance $ds$, the term $L_\textrm{p}$ is overestimating the planet luminosity by a factor of $(1-\frac{R_\textrm{J}}{ds})^{-1}$.  The discrepancy is highest for the closest cells, where $ds$ corresponds to the length of the grid cells.
For our simulations, 
the maximum overestimate would be a factor of only $1.007$, so we neglect this correction for simplicity.  Notice also that the radius of the accreting proto-planet is likely larger than Jupiter's \citep[see e.g.,][]{helled14}.  That can be corrected by choosing a smaller value for the efficiency parameter $\epsilon$.}

\subsection{Radiative Cooling}
The gas cooling term $Q^-$ accounts for the energy radiated by the gas in the vertical direction considering blackbody emission.
We use the same procedure as in \cite{montesinos15}, computing the optical depth as:
\begin{equation}\label{EqTau}
 \tau=\frac{1}{2}\kappa \Sigma.
\end{equation}

Our model assumes a constant opacity of $\kappa=100\,$ cm$^2$/g, consistent with the absorption opacities shown by \cite{kataoka15} for grains of sizes $1\mu \textrm{m}$ to $100\mu \textrm{m}$. We checked that the CPD region is optically thick throughout all the simulations, to ensure that the planet luminosity does interact with the surrounding material.

To model the emission from the disc surface considering its vertical structure we use \cite{hubeny90} prescription for the effective optical depth:
\begin{equation}\label{EqTauEff}
 \tau_{\rm eff}= \frac{\sqrt{3}}{4} + \frac{3\tau}{8} + \frac{1}{4\tau}.
\end{equation}
Then the effective temperature and the cooling term are defined as:
\begin{equation}\label{EqTempEff}
 T_{\rm eff}^4=\frac{T^4}{\tau_{\rm eff}},
\end{equation}
\begin{equation}\label{Cooling}
 Q^- = 2\sigma T_{\rm eff}^4,
\end{equation}
where $\sigma$ is the Stefan--Boltzmann constant and the factor 2 accounts for the upper and lower faces of the disc from which the energy is radiated.

To solve the energy equation including this cooling term we use the implicit solution for the source step, using the same approximations as \cite{commercon11} for the expansion of the power of the temperature.


\subsection{Computational domain, boundary and initial conditions}\label{sec_Comput}


The initial surface density profile of the disc is set as:
\begin{equation}\label{EqInitDens}
 \Sigma(r)= \Sigma_0 \frac{r_\textrm{p}}{r},
\end{equation}
with $\Sigma_0$ the density at the planet location. 
We use $\Sigma_0 = 30 \textrm{g/cm}^2$, which gives a disk mass of $M_\textrm{disk} \sim 10^{-2} M_\odot$ when considering a disk size of 50 AU.

The radial domain goes from 4 AU to 25 AU from the star. At the inner radius we set an open boundary condition, while at the outer radius we impose a constant surface density.


The resolution of our standard simulations is $n_r \times n_s = 320 \times 960$, with $n_r$ and $n_s$ the number of radial and azimuthal sections respectively. We performed two additional simulations with a higher resolution of $n_r \times n_s = 1600 \times 4800$ (a factor 5 increment in each direction), to test the dependence of our results and conclusions with resolution. For the sake of clarity, this comparison is presented as Appendix~\ref{app_hr}.

\subsection{Smoothing Tapers}
In order to smoothly introduce the effect of the planet gravity, the code uses a taper factor, such that the planet mass is ``turned on'' during the first $N_{\rm{taper}}=10$ orbits:
\begin{equation}\label{EqTaper}
 M_\textrm{p}(t) = \begin{cases}
	       10^{-3} M_\star \cdot \sin^2(\frac{\pi}{2}\frac{t}{N_{\rm{taper}}\, T_\textrm{p}})	&	t < 10 T_\textrm{p} \\       
               10^{-3} M_\star      &    	{\rm otherwise}.
      \end{cases}
\end{equation}

Additionally, the system is evolved for 100 orbits before letting the planet accrete, in order to avoid unphysically high accretion rates previous to the gap opening. Then, the accretion rate (eq. \ref{EqAcc}) and the planet luminosity $L_\textrm{p}$ (eq. \ref{EqPLum}) are  multiplied by a taper of the same form as eq.~\ref{EqTaper} between the orbits 100 and 110, to introduce their effect smoothly after the gap formation.

\subsection{Parameter space exploration}

We mostly explore the parameters relevant to the physical processes introduced in this paper, namely, the accretion on to the planet (\S~\ref{sec_AccModel}) and the feedback efficiency (\S~\ref{sec_PlanetHeat}).

For the accretion time-scale of the planet, we use physically plausible values in between the viscous and the free-fall time-scales of the CPD.  These  result on detectable accretion rates, $\simgt 10^{-8} {\textrm{M}_\odot}/{\textrm{yr}}$, and correspond to the formation of a Jupiter-mass planet over a reasonable timescale of $\sim 10^{5} \textrm{yr}$ \citep{montesinos15, zhu15}.
When convenient, we write the accretion timescale in units of the planet orbital time as $1/t_{\rm{acc}} = f_{\rm{acc}}/T_{\rm p}$. 
The accretion fraction $f_{\rm{acc}}$ can be then understood as which fraction of, or how many times, the circumplanetary region is accreted in one orbit of the planet around the star. 

We test different values between 0 and 1 for the efficiency factor $\epsilon$ to take into account that it is not known how much of the energy released by the planet will be absorbed by the gas. 

Most of our models have the planet in a circular orbit.  Since we are interested in exploring the variability of the accretion process, we run two simulations where the planet has an eccentric orbit, with $e=0.1$.

For the effect of varying other physical parameters, such as viscosity, opacity, the masses of the planet, disc and star, or the planet semi-major axis, we refer the reader to \cite{montesinos15}.
A resolution test is presented in the Appendix.



\section{Accretion rate and its variability} \label{sec_AccretionVariability}

%
%

Arguably, the most important quantities in our study are the planet accretion rate and  luminosity, which can be measured by observations, potentially revealing the presence of a planet.

\subsection{Average accretion rates}
\label{sec:avrgacc}

\begin{figure}
\centering
\includegraphics[width=84mm]{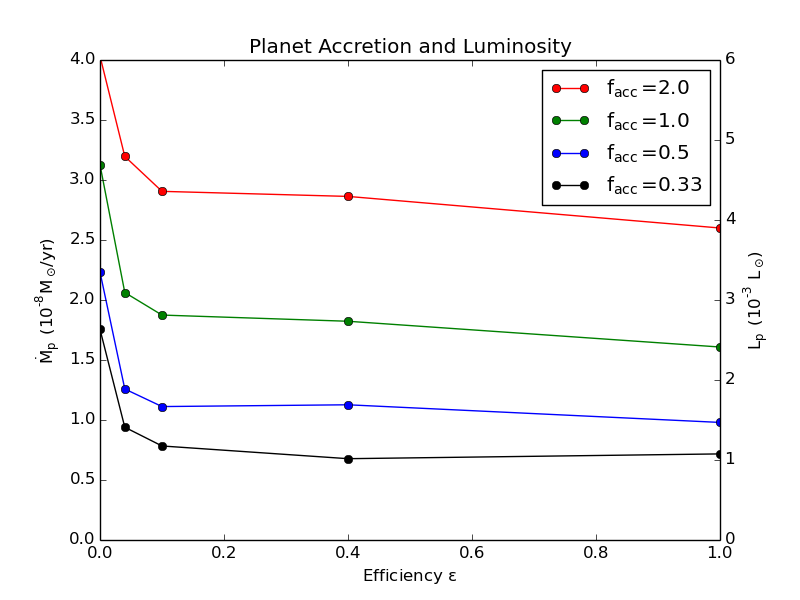}
 \caption[Planet accretion rate and luminosity vs. the feedback efficiency $\epsilon$.]{Planet accretion rate (left axis) and luminosity (right axis) as a function of the feedback efficiency $\epsilon$, for different values of $f_\textrm{acc}$, for planets in circular orbit. The plot shows the monotonic decrease of $\dot{M}_\textrm{p}$ with $\epsilon$. Most of the variation occurs at low efficiencies, and stays roughly constant for $\epsilon \simgt 0.1$.}
  \label{FigFull_Acc2}
\end{figure}

Figure~\ref{FigFull_Acc2} shows the average accretion rate for simulations of a planet in a circular orbit, in which the parameters controlling the accretion timescale and the feedback efficiency are explored.  Averages are taken from the last 100 snapshots, separated by an interval of half the planet orbit, between the orbits 300 and 350.
Even though $\epsilon$ varies from 0 to 1, and the accretion timescale varies within a factor 6, the accretion rate is confined to a relatively narrow range between 0.7 and 4 $\cdot 10^{-8} \textrm{M}_\odot/\textrm{yr}$, with the planet luminosity remaining in the range 1--6 $\cdot 10^{-3} \textrm{L}_\odot$.  As analysed in detail by \cite{montesinos15}, such planet luminosities could easily reproduce the ``hot spots'' observed in several proto-planetary discs.

As expected, for larger accretion fractions $f_\textrm{acc}$ (shorter accretion timescales $t_\textrm{acc}$) $\dot{M}_\textrm{p}$ increases by a factor of a few. However, the main feature of the plot is the drop in the accretion rate when the planet feedback is turned on.
The drop in accretion is due to a lower density of the CPD region.
Feedback heats up preferentially the inner region of the CPD, which produces a steeper pressure gradient, thus a greater force that pushes the material outside the CPD. The pressure gradient continues to keep the material from reentering the CPD.
This process was already discussed by \cite{montesinos15}, showing that the heating alone can stop the material from accumulating near the planet, even when accretion is not considered.

As seen in Fig.~\ref{FigFull_Acc2}, feedback causes a decrease of 35\%--50\% in the accretion rate compared to the fiducial case when no feedback is considered. 
Interestingly, most of this effect takes place at efficiencies as low as $\epsilon \leq 0.1$, and remains approximately constant from there to $\epsilon=1.0$.  
This is relevant because even if the feedback is not 100\% efficient, or if the calculation of the planet luminosity is overestimated by any reasons, the feedback would still be effective, dropping the accretion rate and luminosity to a well-determined value in our models.  

Given the results above, and the fact that  \cite{marleau17} showed that the energy released from the accretion on to the planet corresponds to $\epsilon \sim 1$, in the following we concentrate on models with $f_\textrm{acc} = 1$ and $\epsilon=0$ and 1 only.


\subsection{Feedback variability and its origin}

Another interesting quantity from the observational point of view is the variation of the planet accretion and luminosity with time. If the variations are too intense, then one-time observed planet signatures might not be good indicators of the planet properties.  However, characterising the variability may reveal the physical processes at work.




\begin{figure}
\centering
\includegraphics[width=84mm]{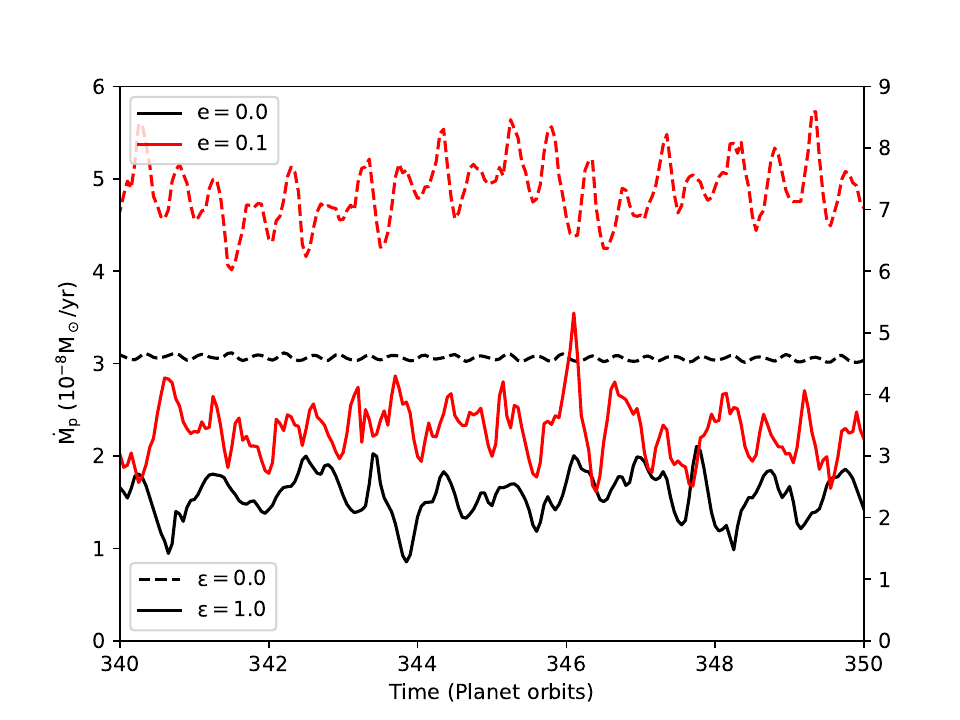}
 \caption[Planet accretion rate and luminosity evolution. Eccentricity efficiency study.]{
 Time evolution of the planet accretion rate (left axis) and luminosity (right axis), for the cases of planets in circular (black lines, $e=0$) and eccentric orbits (red lines, $e=0.1$),  without feedback (dashed lines, $\epsilon=0$) and with feedback (solid lines, $\epsilon=1$), from the orbits 340 to 350.
 }
  \label{Fig_Variability_Ecc}
\end{figure}

\begin{figure}
\centering
\includegraphics[width=84mm]{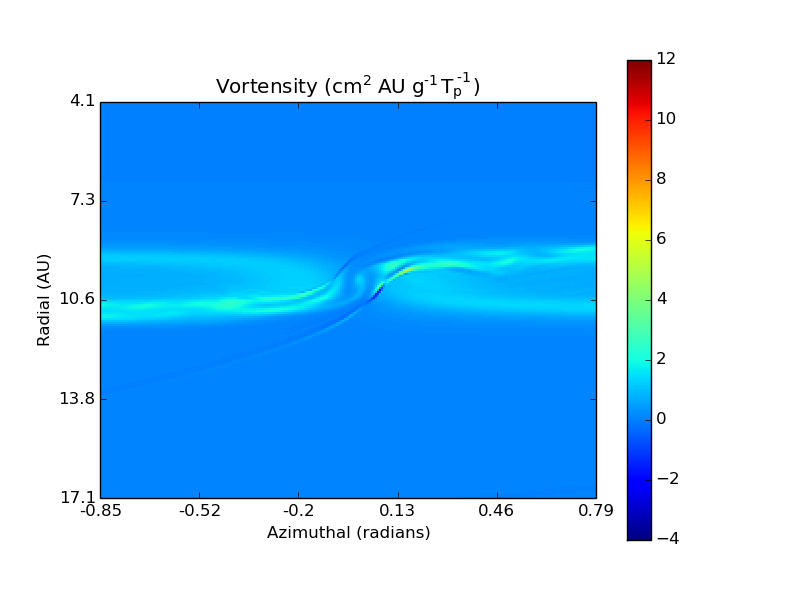}
\includegraphics[width=84mm]{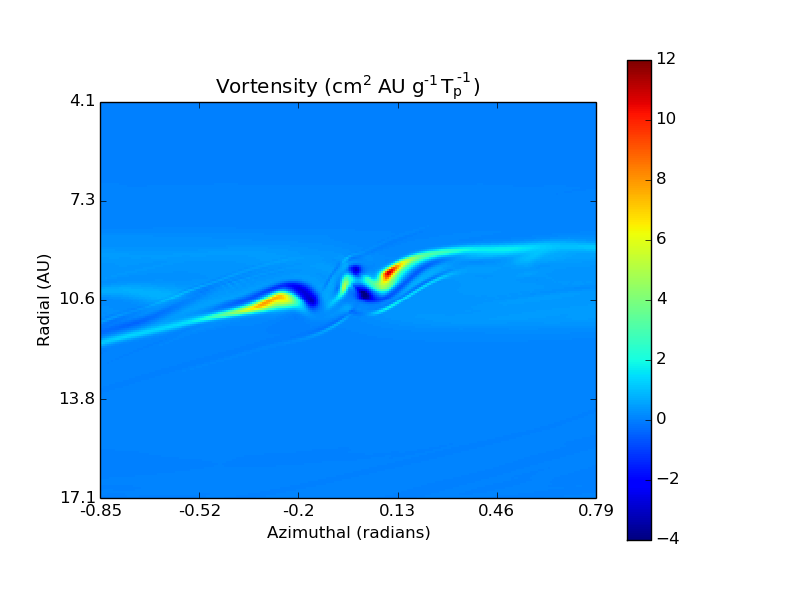}
 \caption[Vortensity maps.]{Vortensity maps for the simulations of planets in circular orbit, for the cases without feedback ($\epsilon=0$, top) and with feedback ($\epsilon=1$, bottom), at a time of 350 orbits. We can notice vortex like structures in the inner and outer edges of the gap when the feedback is considered.}
  \label{FigColor_Vortensity}
\end{figure}

We start analyzing the case of a planet in circular orbit ($e=0$).
Figure \ref{Fig_Variability_Ecc} shows in black lines the planet accretion rate and luminosity variability over time. 
We notice that the variability of the accretion rate increases by an order of magnitude when the feedback effect is considered, reaching a standard deviation $\Delta \dot{M}_p \sim 0.1 \dot{M}_p$.

To explore the physical origin of the variability, we create vortensity maps, defined as:
\begin{equation} \label{eq_vortensity}
\overrightarrow{\omega} = \frac{\nabla \times \overrightarrow{v}}{\rho}.
\end{equation}

Figure \ref{FigColor_Vortensity} shows that the feedback luminosity induces vortex like structures close to the circumplanetary disk, that then travel along the coorbital region. These vortices are short lived, and disappear within an orbital timescale.
We infer that the feedback luminosity of the planet excites the gas around the circumplanetary region, creates perturbations through the coorbital region, and that these in turn contribute to increasing the accretion variability.


%
%
%
\subsection{Power spectra for different planet parameters}

Besides feedback, other physical effects can increase the variability of the accretion on to a planet.  Here we test the effect of orbital eccentricity.  As the planet traverse the disc on an eccentric orbit, it will reach the higher density regions at the edges of the gap, increasing its accretion rate at peri- and apocentre.

Figure \ref{Fig_Variability_Ecc} shows the accretion variability of planets in circular and eccentric ($e = 0.1$) orbits. 
In terms of amplitude, the variability of a planet in an  eccentric orbit, with or without feedback, is similar to that of a planet in circular orbit considering the feedback effect. In order to distinguish between these three scenarios, we characterise the time variability using power spectra {to isolate the periodic features from the rest of the variability, and measure their characteristic timescales.}

{We used a variant of the method described in \citet{arevalo2012} to construct power spectra estimates of the accretion time series. This method first convolves the time series with a sinc function, sinc$((t-t_i)\times2\pi/\tau_1)$ to remove fluctuations on timescales shorter than $\tau_1$ and then with a sinc$((t-t_i)\times2\pi/\tau_2)$ to remove fluctuations on timescales shorter than $\tau_2$. The difference of the convolved light curves then retains only variations on timescales between $\tau_1=\tau\times\sqrt{(1+\Delta)}$ and $\tau_2=\tau/\sqrt{(1+\Delta)}$, where we used a value of $\Delta=0.01$. The normalized variance of the filtered and subtracted time series, $\dot{M}_\tau(t_i)$, is calculated as:}
\begin{equation}
\sigma^2_{\rm norm,\tau}=\frac{\sum_{i=1}^N (\dot{M}_\tau(t_i)-<\dot{M_\tau}>)^2}{N<\dot{M_\tau}>^2}.
\end{equation}
{This value is then divided by the frequency range ($1/\tau_2-1/\tau_1$) to produce an estimate of the dimensionless variability power at timescale $\tau$. The process is repeated for different values of the timescale to construct the power spectrum. This method reduces considerably the effects of red-noise leak in non-stationary time series, where strong trends underlie the variations of interest, as is the case in some of the segments studied. The method follows the general prescription of \citet{arevalo2012} but changes the convolution function from Gaussian to sinc, allowing a better frequency resolution of the power spectrum estimate. }

\begin{figure}
\centering
\includegraphics[width=90mm]{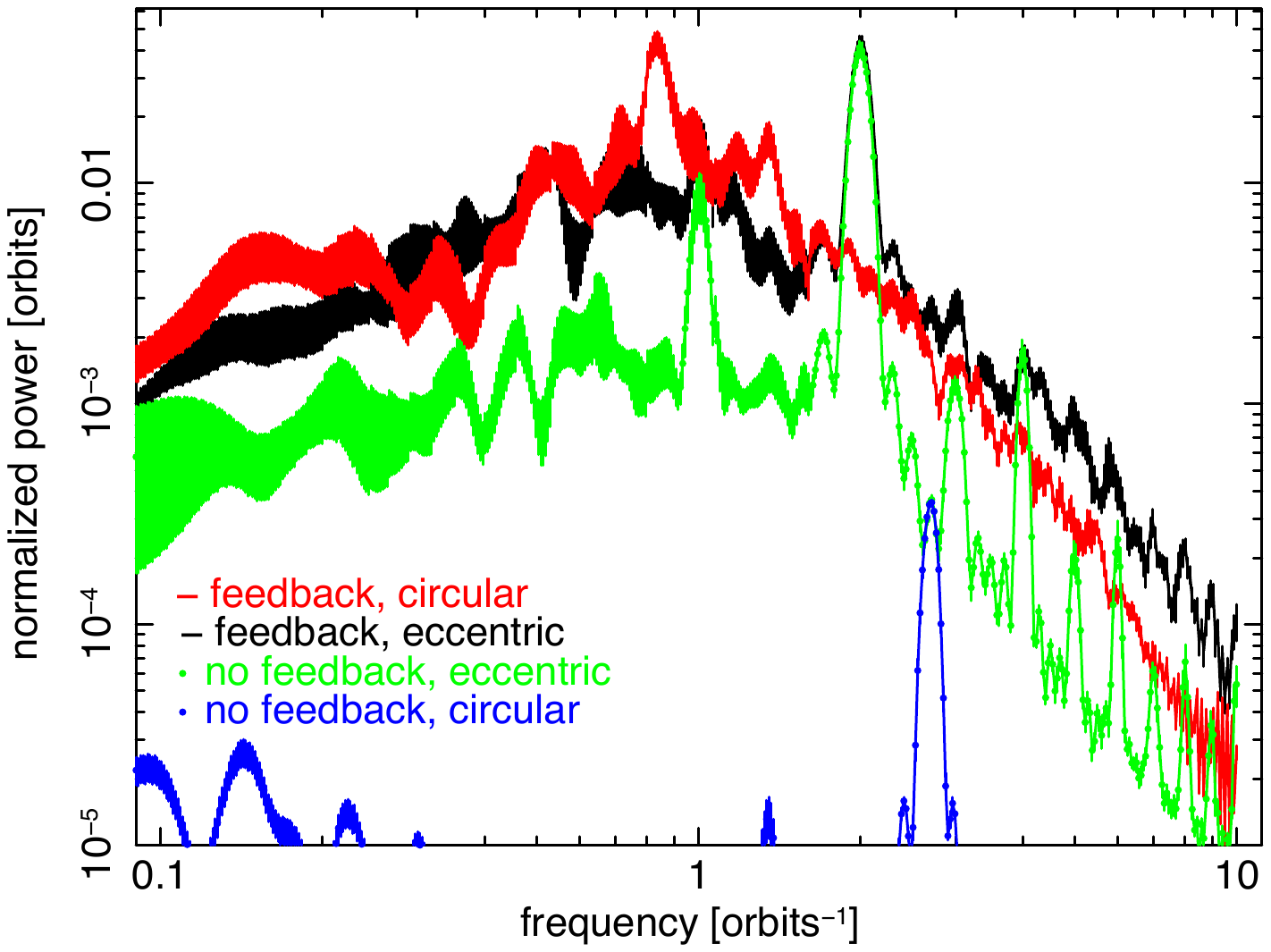}
 \caption[Power spectra]{Power spectra of the accretion rate light curves normalized by average accretion rate. Feedback results in broadband variability power, peaking at super-orbital timescales, while eccentricity imprints an additional strong peak at half-orbital timescales, a weaker peak at orbital timescales, and higher harmonics of both.}
  \label{Fig_PDS_log}
\end{figure}

Figure~\ref{Fig_PDS_log} shows the power spectra of accretion rate time series of the four cases: circular and eccentric planet orbits, each with and without feedback. Each curve shown is the average of 10 power spectra calculated from consecutive segments, each 12.8 orbits long, starting from orbit 200. The integral of the plotted curves equates to the normalized variance $\sigma^2$ on timescales between 0.1 and 12.8 orbits. 


The eccentric orbit cases (black and green in Fig.~\ref{Fig_PDS_log}) have a clear, large amplitude peak at a frequency of 2 orbits$^{-1}$, or equivalently a period of 0.5 orbits, with a smaller peak at a timescale of 1 orbit, and higher frequency harmonics of both. Using the error on the mean of the 10 segments in each power spectrum as error bars, a Gaussian fit to the main peak results {in almost identical peak frequencies,  $1.998\pm{0.001}$ ($1\sigma$ interval) for the case without feedback and $2.001\pm{0.002}$ for the case with feedback.} 

The normalized variance of this periodic signal at a timescale of 0.5 orbits is almost exactly the same for the case with and without feedback, $\sigma^2=0.00824$  and  $\sigma^2=0.00807$, respectively, i.e. within 3\% of each other. The width of the Gaussian fits is consistent with the width expected from a pure sinusoid time series subjected to the same analysis.   

The main difference introduced by feedback is the stronger broadband variability power. In all cases with either feedback or eccentricity (or both), the power spectrum broadly resembles a broken powerlaw shape. Below the break the slope is approximately 0, consistent with white noise, while above the break the powerlaw slope ranges from -2 to -3, showing a steep and steady decline of variability amplitude with increasing frequency. The break occurs around the orbital timescale for the cases with feedback and around half this timescale for the case with eccentricity and no feedback. 

The variance of the broadband noise is enhanced by the feedback effect: integrating the power spectra between timescales of 0.1 and 12.8 orbits, the time series with feedback have normalized variances of $\sigma^2=0.027$ (eccentric) and $\sigma^2=0.024$ (circular) while the broadband noise of the eccentric case without feedback only amounts to 
$\sigma^2=0.003$. Evidently, the time series with neither eccentricity nor feedback shows orders of magnitude smaller variance.

Since the power in the periodic signal of the eccentric time series is independent of feedback while feedback enhances the broadband noise, the contrast between periodic and aperiodic variability powers can potentially be used as a diagnostic for feedback strength. For the particular setups used in these simulations, the 0.1--12.8 orbit variance is composed of 75\% periodic signals and 25\% broadband noise, when no feedback is included. Therefore, for this planet mass and eccentricity, the periodic signal can at most contribute 75\% of the variance. Adding feedback increases the contribution of the broadband noise; for the feedback strength used here the variance is composed of 33\% periodic signals and 67\% broadband noise, i.e., the latter becomes dominant. 

\section{Feedback effect on the gap} \label{sec_HeatingGap}

Although the gap is not directly involved in our numerical implementation of accretion feedback, it can be affected by this process through the transport of mass and energy from the CPD region. In this section we will show the different effects of the planetary accretion and feedback in the gap properties. 
We show results for planets in circular orbit, varying the accretion and feedback parameters ($f_\textrm{acc}$, $\epsilon$).

\begin{figure}
\centering
\includegraphics[width=84mm]{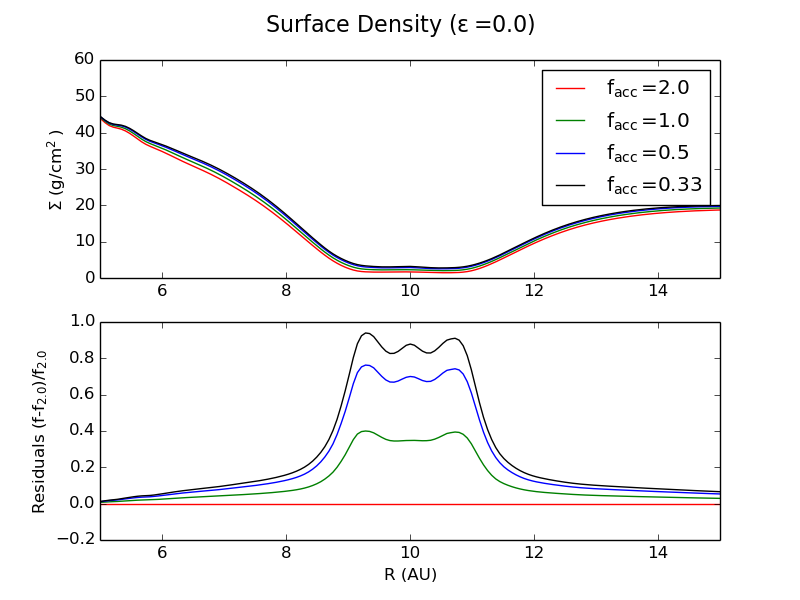}
 \caption[Azimuthally averaged density profiles for discs with $\epsilon=0.0$.]{Top: Azimuthally averaged density profiles for discs with $\epsilon=0.0$ and different accretion time-scales.
 Bottom: Residuals normalized respect to the $f_\textrm{acc}=2.0$ curve of the top panel.
 We can see how the overall density in the gap is reduced for higher accretion fractions, since more mass is removed from the simulation.}
  \label{FigFull_ProfileDensAcc}
\end{figure}

The effect of accretion on the density profile was already studied by \cite{duermann15}, using an equivalent prescription for the accretion rate. In their results the accretion deepened the gap carved by a planet. Our azimuthally-averaged density profiles in Figure \ref{FigFull_ProfileDensAcc} show the same behaviour 
-- for higher accretion fractions $f_\textrm{acc}$ more mass is removed, deepening the gap. 
In order to directly compare different models, we do not add the accreted material to the planet mass.  If that was the case, higher $f_\textrm{acc}$ would produce more massive planets, which in turn would carve even deeper gaps.

\begin{figure}
\centering
\includegraphics[width=84mm]{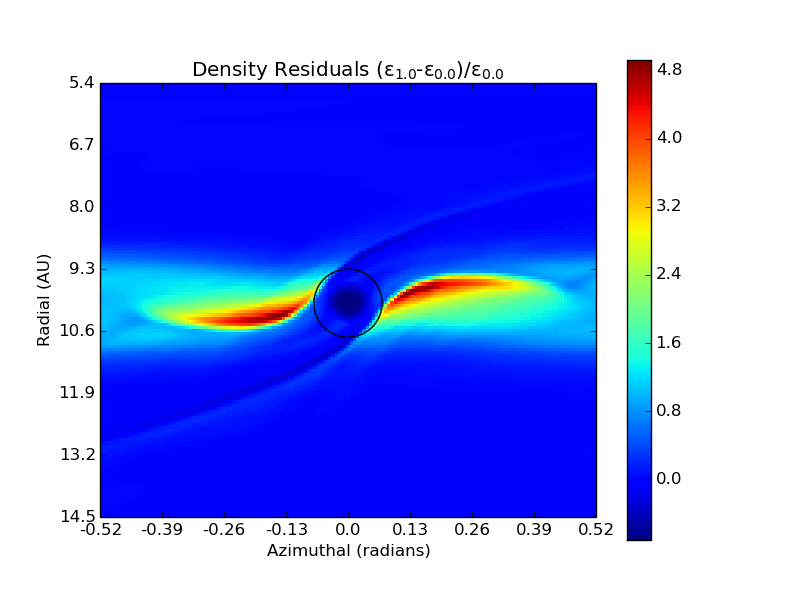}
 \caption[Surface density comparison between the case $\epsilon=1.0$ with $\epsilon=0.0$.]{Surface density comparison between the case $\epsilon=1.0$ and $\epsilon=0.0$, using the residual $(\Sigma(\epsilon_{1}) -\Sigma(\epsilon_{0}))/\Sigma(\epsilon_{0})$.
 Both simulations have  $f_\textrm{acc}=1.0$. 
 The plot shows that the simulation with $\epsilon=1.0$ has less mass in the CPD, and more mass in the gap region than the case with $\epsilon=0.0$. 
The Hill radius around the planet is marked with a black circle.}
  \label{FigColor_Dens}
\end{figure}

\begin{figure}
\centering
\includegraphics[width=84mm]{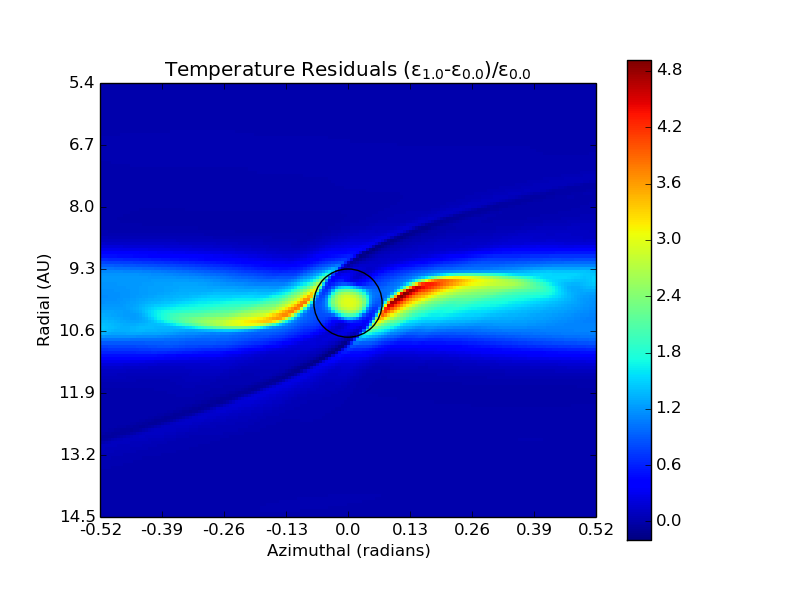}
 \caption[Temperature comparison between the case $\epsilon=1.0$ with $\epsilon=0.0$.]{ Same as Fig.~\ref{FigColor_Dens}, but showing the temperature residual, $(T(\epsilon_{1}) -T(\epsilon_{0}))/T(\epsilon_{0})$.
   The plot shows that the simulation with $\epsilon=1.0$ reaches higher temperatures both in the CPD region and inside the gap. } 
  \label{FigColor_Temp}
\end{figure}

To study the effect of feedback we present a comparison of the density and temperature distributions between the cases with $\epsilon=1$ and $\epsilon=0$ (Figures \ref{FigColor_Dens} and \ref{FigColor_Temp}), using the residuals $(\Sigma(\epsilon_{1}) -\Sigma(\epsilon_{0}))/\Sigma(\epsilon_{0})$, and $(T(\epsilon_{1}) -T(\epsilon_{0}))/T(\epsilon_{0})$. 
The temperature increases when feedback is on, both in the CPD and the gap, as expected.  
Density, in contrast, decreases in the CPD with feedback while it increases in the gap.  This behaviour can be understood as due to the increase of pressure in the region closer to the planet, as discussed in \S~\ref{sec:avrgacc},
and was already observed in \cite{szulagyi17_Feedback}'s 3D simulations when switching the planet temperature.

\begin{figure}
\centering
\includegraphics[width=84mm]{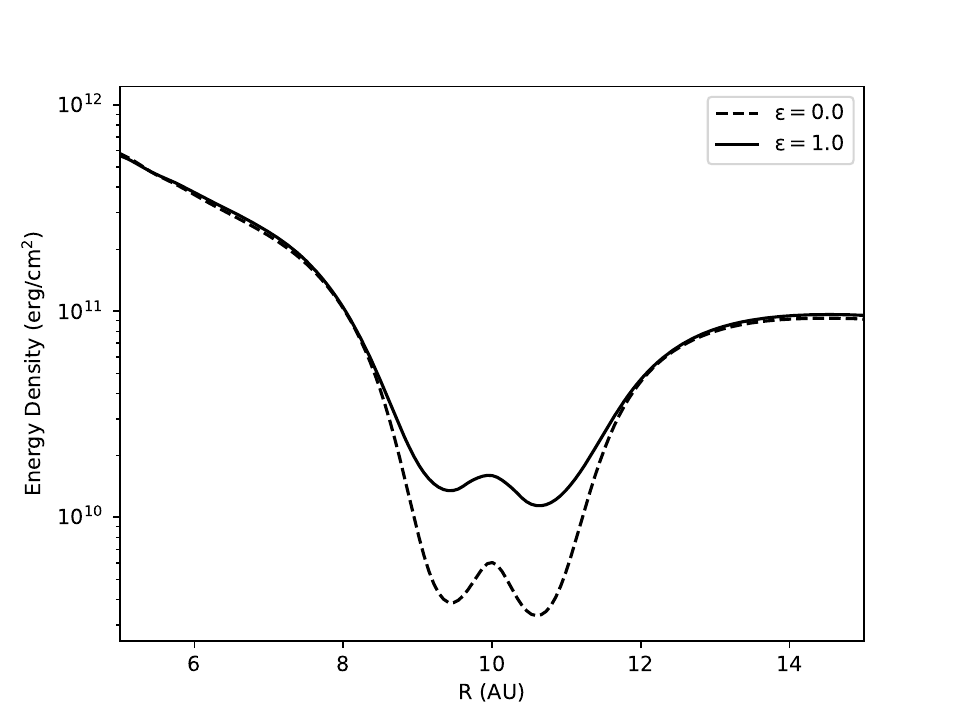}
 \caption[Azimuthally averaged energy profiles.]{
 Azimuthally averaged energy surface density profiles. The plots shows the cases with and without feedback (solid and dashed lines respectively), for a planet in circular orbit.}
  \label{Fig_Profile_Energy}
\end{figure}
\begin{figure}
\centering
\includegraphics[width=84mm]{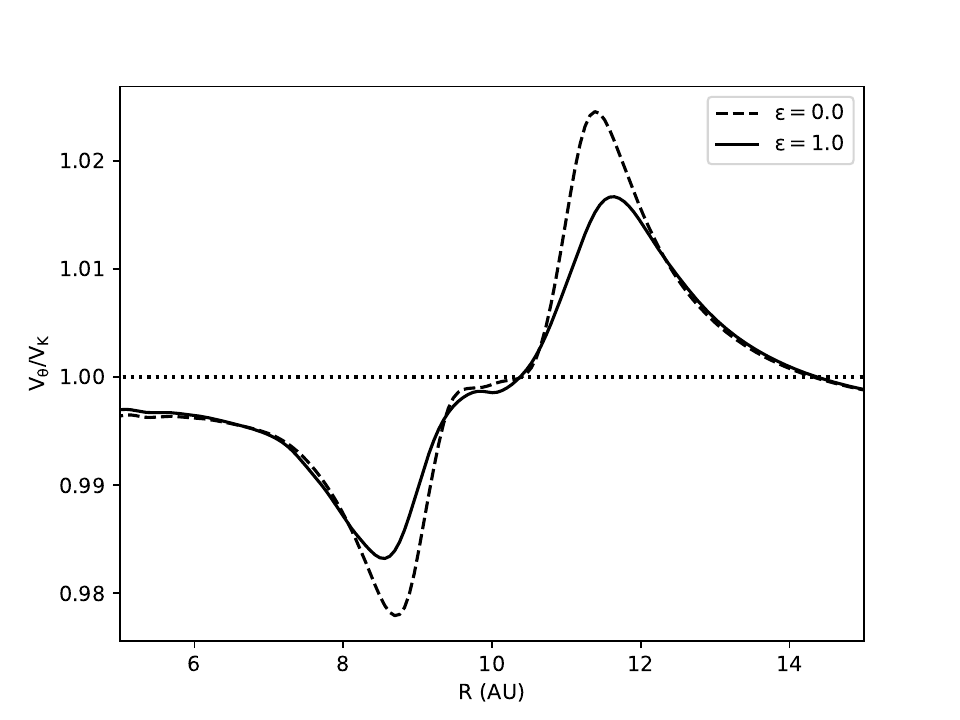}
 \caption[Azimuthal velocity profile]{
 Azimuthally averaged orbital velocity profile (normalized to the Keplerian value). The plots shows the cases with and without feedback (solid and dashed lines respectively), for a planet in circular orbit.}
  \label{Fig_Profile_AzimithalVel}
\end{figure}

Combining the results for density and temperature we can obtain the energy density field (eq. \ref{EqIdealEnergy}) of the disc. 
Figure \ref{Fig_Profile_Energy} shows that the energy surface density in the gap increases by effect of the planet luminosity,  
as expected. 
The gas pressure is proportional to the energy surface density (eqs. \ref{EqIdealGas} and  \ref{EqIdealEnergy}), therefore the gas pressure in the gap increases 
and this results in a lower pressure gradient at the gap edges.

The radial pressure gradient directly affects the orbital speed of the gas, changing its value from the standard Keplerian profile.
The effect can be directly seen in the azimuthal velocity profiles in Figure \ref{Fig_Profile_AzimithalVel}, which take values closer to  Keplerian when the feedback is on.
Although the azimuthal velocity differs from Keplerian only by a 2\% for $\epsilon=0.0$, and 1\% for $\epsilon=1.0$, this contribution is very significant compared to that of the gas pressure alone in a disc with no planet, where the variation is just 0.1--0.2\%, as can be seen far from the planet potential in the same plot.  
A change in the orbital velocity has a direct impact on the dust drift, as will be discussed in \S~\ref{sec_Discuss_DustConcentration}. 



\section{Discussion}\label{sec_Discuss}

Our simulations showed the diverse effects of feedback from an accreting planet on the CPD and the coorbital region. The most relevant results are that Jupiter-like planets can still achieve detectable accretion rates, {present high broadband-variability,} and that their heating affects the whole gap, not only the CPD region.

\subsection{Feedback as a Slow Down Mechanism for Accretion}
In the core accretion model \citep{pollack96}, Jupiter-mass planets can enter into a runaway accretion phase, being able to double their masses on timescales as short as $10^4\,$yr. %
In the disc fragmentation scenario \citep{boss97}, planets typically also grow very quickly, reaching the brown dwarf regime \citep[][but see also \cite{nayakshin15}]{stamatellos15}.

It is interesting to note that in our model feedback acts as a limiting mechanism for planet growth, decreasing the accretion rate in 35\%--50\% (see Fig.~\ref{FigFull_Acc2}), but suggesting that accretion can still proceed {(cf. Sec.~\ref{sec_Discuss_Limitations} below).}
%

\subsubsection{Are high feedback efficiencies valid?}
\label{sec_Discuss_Feedback}

This question is crucial for the validity of our model, since our results depend on the planet accretion energy being actually absorbed by the CPD.  Still, an efficiency factor as low as $\sim 10\%$ is enough for the feedback to have a strong effect (see Fig.~\ref{FigFull_Acc2}).

Obtaining a realistic value for the efficiency would require a more sophisticated numerical approach, including a treatment of radiative transfer. Still, we can provide two simple arguments to justify that the radiation of the planet should be absorbed at least partially by the gas.  
The first one is the geometry of the CPD; \cite{szulagyi16} has shown that the scale height of the disc increases for high feedback values, even to the point of turning the disc into an envelope. Our own 2D models show that the temperature reaches its maximum at the location of the planet.  Therefore, we can expect that most of the radiation will go through the gas. 
Additionally, whether the radiation is absorbed depends on the optical thickness of the gas. In our models we found a high enough optical depth, as was already seen by \cite{ayliffe09a,ayliffe09b}, and confirmed in the simulations of \cite{szulagyi16}, which use the flux-limited diffusion approach. 

Finally, the simulations of \cite{marleau17} also confirm that most of the accretion energy available is transported back into the CPD.
From these previous studies we are confident that the feedback efficiency $\epsilon$ should have values large enough in order for feedback to be relevant.

\subsection{An eccentric planet or an accreting planet?} \label{sec_Discuss_Eccentricity}

We have shown that feedback makes the accretion on to the planet, and therefore its luminosity, variable at a $\sim10$\% level, which can allow us to identify this process in observations over several epochs.  However, there are other mechanisms that could cause the planet luminosity to vary, such as it being on an eccentric orbit.  \cite{dunhill15} showed that a planet on a low-eccentricity orbit ($e=0.1$) produces a variable accretion rate, which is confirmed by our simulations.

Fortunately, the characteristics of the variability can be used to disentangle between both cases.  While a planet on an eccentric orbit produces a periodic signal, the effect of feedback is seen as a broadband noise.  Observing the system for several orbital periods would then reveal which of both processes are relevant.  While our nominal system orbital period of $\approx 30\,$yr makes this long-term observation unfeasible, the results should also hold for planets in short-period orbits, such as the precursors of the large known population of hot Jupiters.
We note that the Planet Formation Imager (PFI) project aims to resolve the gap of a Jupiter-mass planet at 1 au from its host star, allowing this kind of monitoring to be performed during the next decade \citep{monnier19}.

\subsection{Decrease in the dust concentration} \label{sec_Discuss_DustConcentration}

Besides the effect of the feedback on the gas, which we have modelled here, we can also expect the dust to be affected by the changes in the gas dynamics. On top of the gravitational potential, gas also feels pressure forces, which makes its orbit slightly non-Keplerian.
If the gas is sub-Keplerian, then the dust grains will lose angular momentum due to friction with the gas, and drift inwards as a result. In the opposite case, if the gas is super-Keplerian, dust will drift outwards. In terms of the pressure, this can be seen as the dust concentrating at the pressure maxima \citep{weidenschilling77, pinilla12}.

The analysis of the gas orbital speed  (Fig.~\ref{Fig_Profile_AzimithalVel}) let us infer how the dust grains behave. 
Our simulations show that when the feedback is turned on, the pressure gradient is smoother and the orbital velocity becomes more Keplerian.  
At the inner edge of the gap, where the pressure gradient is negative, dust drift should be inwards, but slower than in the case with a planet without feedback. The outer edge is more interesting, since here the gas is super-Keplerian, which would allow for dust particles to drift outwards and concentrate  at the local pressure maximum outside the carved gap \citep{rice06, fouchet07}. 
Our results however, show that this concentration should be less efficient, and therefore reduce the expected contrast in observations.
A proper quantification of this effect is out of the scope of this paper, but should be a promising topic for future studies.

\subsection{Model limitations}
\label{sec_Discuss_Limitations}

The weakest point of our model is the lack of the vertical coordinate. The recent 3D studies of \cite{szulagyi14, szulagyi16} have shown the accretion from the CPD onto the planet has a large meridional component caused by the enhancement of the CPD scale height.
With our 2D model (and assuming vertical hydrostatic equilibrium) we also see that the scale height increases with the planet luminosity.
Further 3D effects, such as the formation of a circumplanetary envelope \citep{szulagyi16, szulagyi17_Feedback}, cannot be followed with our approach.  %
In any case, our 2D approach should be adequate to characterise the changes in the co-orbital region, including the gap edges, at a qualitative level.  Therefore, we expect the feedback-induced variability and the shallower pressure gradient to appear also in 3-dimensional models.

Additionally, with a global disc model we cannot afford to resolve the hydrodynamics down to the planetary surface, so cannot really follow the accretion properly.  We take the common approach of mimicking the accretion with a recipe that depletes the CPD region on a given time-scale \citep[e.g.,][]{kley99, zhu11, tanigawa16, duermann17}, so the accretion rate is roughly proportional to the CPD mass.

We treat that time-scale as a free parameter, but confined to a physically motivated value, namely in between the dynamical and the viscous time-scales.
Our results show that the accretion rate depends on that choice, as expected.  However, the variation is confined within a factor of a few only, and all our conclusions remain valid.

{Due to the relatively low resolution, the material is depleted from within a radius $0.45 R_{\rm{Hill}}$ (eq.~\ref{EqAcc}).  At the same time, the feedback is directly implemented within a radius $\sim 0.5 R_{\rm{CPD}} = 0.3 R_{\rm{Hill}}$, which is motivated by the large optical depth and geometrical thickness of the CPD (see Sec.~\ref{sec_Discuss_Feedback} above).
This near coincidence is unfortunate, as freshly-heated material might be too quickly accreted, before it can have a stronger effect.  Nevertheless, we do see a factor-two heating at $\sim R_{\rm{Hill}}$ in our models (Fig.~\ref{FigColor_Temp}), consistent with what \cite{ayliffe09a} found (see their Fig.~4). 
In any case, simulations with higher resolution and separate scales for both processes, i.e., a smaller accretion radius and a larger heating radius, would be needed to confirm our results.}

Previous works have studied the effect of radiative feedback on the migration of the forming planet \citep{nayakshin13, stamatellos15, benitez15}.  In our models the planet radius was kept fixed, so migration cannot be directly measured from the simulations.  A post-processing measurement of the torque acting on the planet was unfortunately inconclusive.  The torque magnitude, and even its sign, depended very strongly on how the CPD was defined \cite[see][]{crida09}.  As this is the region where both accretion and feedback act directly, our present numerical set up is simply not adequate to address this issue and we defer it to a follow-up work.


\section{Summary}\label{sec_Summary}
The accretion on to giant planets and their associated luminosity can produce observational signatures that reveal the process of planet formation. In this work we study how the accretion and luminosity of a gas giant affects the circumplanetary disc, and the surrounding gap. We extended the procedure used in \cite{montesinos15} by taking into account the accretion on to the planet, and computing the planet luminosity accordingly. 
Our simulations show that planet accretion feedback decreases the accretion rate by at most a 50\%, suggesting that detectable accretion luminosities can develop{, although further studies that accurately resolve the accretion and heating in the CPD region are still required to verify this result.}


Interestingly, the feedback perturbs the dynamics of the gas surrounding the planet, creating vortices and making the accretion rate on to the planet vary stochastically.  Such variation pattern is very different to the one produced by a planet in an eccentric orbit, which is characterised by a well-defined periodicity.  Monitoring over several orbits would therefore allow to disentangle which or both processes are relevant once a proto-planet candidate is identified.

The feedback effects are not limited to the CPD -- the gap carved by the planet is partially heated up and refilled, making the pressure profile smoother. 
As a consequence, the orbital velocity in the gap becomes more Keplerian. This is likely to influence the dust transport and evolution by decreasing the rate of concentration of particles at the dust traps.

To conclude, understanding planet accretion feedback is both important to correctly interpret the accretion signatures of proto-planet candidates, and to properly model the gas and dust dynamics in the CPD and gap regions.

\section*{Acknowledgments}
We would like to thank P.\ Ben\'itez-Llambay, W.\ Lyra, H.\ Klahr, C.P.\ Dullemond, N.\ Cuello, S.\ Stammler, D.\ Stamatellos, S.\ Casassus, J.A.\ Eisner, and G. Marleau for their comments, suggestions and helpful discussions that greatly improved the quality of this work.
We would also like to acknowledge {Sergei Nayakshin and} the anonymous referees who reviewed a previous version of this work.
This work was partly carried out while JC was on sabbatical leave at MPE.  JC and MG acknowledge the kind hospitality of MPE, and funding from the Max Planck Society through a ``Partner Group'' grant. We acknowledge financial support from the ICM (Iniciativa Cient\'ifica Milenio, Chilean Ministry of Economy) via grant RC130007 (MG, JC \& MM) and the N\'ucleo Milenio de Formaci\'on Planetaria grant (JC \& MM),
from CONICYT-Chile through FONDECYT (1141175, JC \& MG; 1150345, MG), and Basal (PFB0609, JC \& MG) grants. 
MG acknowledges financial support from the European  Research  Council  (ERC)  under  the  European  Union's Horizon 2020 research and innovation programme (grant agreement No 714769), from the Deutsche Forschungsgemeinschaft (DFG, German Research Foundation) Ref no. FOR 2634/1, from the Deutsche Forschungsgemeinschaft (DFG, German Research Foundation) under Germany's Excellence Strategy – EXC-2094 – 390783311, and from the Alexander von Humboldt Foundation in the framework of the Sofja Kovalevskaja Award endowed by the Federal Ministry of Education and Research.
MM acknowledges financial support from the Chinese Academy of Sciences (CAS) through a CAS-CONICYT Postdoctoral Fellowship administered by the CAS South America Center for Astronomy (CASSACA) in Santiago, Chile.
PA acknowledges financial support from CONICYT PIA ACT172033 and the Max-Planck Society through a Partner Group grant with MPA. 

\section*{Data availability}

The data underlying this article will be shared on reasonable request to the first author.



\appendix
\section{High resolution comparison}  \label{app_hr}
\begin{figure}
\centering
\includegraphics[width=84mm]{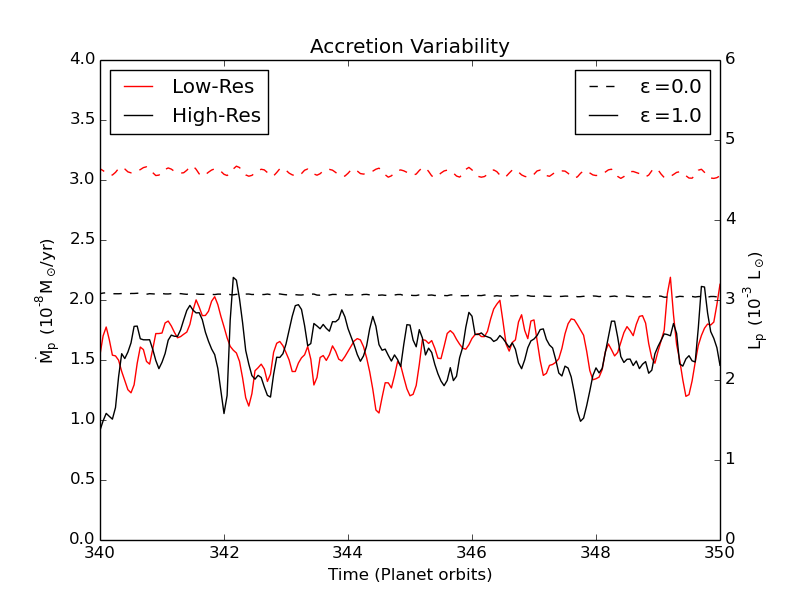}
 \caption[Planet accretion rate and luminosity vs. time. Resolution Comparison.]{Same as Figure \ref{Fig_Variability_Ecc} for the circular cases with high and low resolution. The cases with the feedback turned on present a similar accretion rate and variability in both resolutions.}
  \label{FigHR1_Acc}
\end{figure}

Here we compare the high resolution simulations described in \S~\ref{sec_Comput} with their standard resolution counterparts in terms of accretion rate and variability. 
In Figure \ref{FigHR1_Acc} we observe that the results with feedback are numerically converged, with a very similar accretion rate and variability regardless of resolution.  The cases without feedback however show a $\sim30$\% lower accretion rate for the run with five times higher resolution.  We conclude that the fact that feedback decreases the accretion rate and increases its variability is robust to resolution, although computing the actual rate within the framework of our model in the case without feedback would require even higher resolution simulations.

\bsp 
\label{lastpage}

\end{document}